# Short-Time Dynamics of the Three-Dimensional Fully Frustrated Ising Model

V. A. Mutailamov [a],* and A. K. Murtazaev [a, b]

[a] Institute of Physics, Dagestan Scientific Center, Russian Academy of Sciences,
ul. Yaragskogo 94, Makhachkala, 367003 Russia

[b] Dagestan State University, ul. M. Gadzhieva 43a, Makhachkala, 367025 Russia

*e-mail: vadim.mut@mail.ru

The critical relaxation from the low-temperature ordered state of the three-dimensional fully frustrated Ising model on a simple cubic lattice has been studied using the short-time dynamics method. Particles with the periodic boundary conditions containing N = 262144 spins have been studied. Calculations have been performed by the standard Metropolis Monte Carlo algorithm. The static critical exponents of the magnetization and correlation radius have been obtained. The dynamic critical exponent of the model under study has been calculated.

The study of dynamic critical properties of spin systems is one of the topical problems of modern statistical physics and phase transition physics. Theoretical and experimental studies have provided considerable successes in this field. Nevertheless, the construction of the strict and consistent theory of dynamic critical phenomena on the basis of microscopic Hamiltonians is one of the central problems of the modern theory of phase transitions and critical phenomena, which is still far from its solution.

Recently, for studying the critical dynamics of models of magnetic materials, physicists started successfully using the short-time dynamics method [1–5], in which the critical relaxation of the magnetic model from the nonequilibrium state to the equilibrium one is studied within the model A (the Hohenberg–Halperin classification of universality classes of the dynamic critical behavior [6]). It is considered traditionally that the universal scaling behavior takes place only in the thermodynamic equilibrium state. However, it was shown that this behavior in some dynamic systems can be observed at the early stages of their time evolution from the high-temperature disordered state to the state corresponding to the phase transition temperature [7]. Such a behavior occurs after some time interval which is large enough in the microscopic sense but remains small in the macroscopic one. An analogous picture is also observed in the case of the evolution of the system from the lowtemperature ordered state [1, 2].

Using the renormalization group method, the authors of [7] showed that the kth moment of the magnetization far from the equilibrium point after a microscopically small time interval has the scaling form

$$M^{(k)}(t,\tau,L,m_0) = b^{-k\beta/\nu}M^{(k)}(b^{-z}t, b^{1/\nu}\tau, b^{-1}L, b^{x_0}m_0), \qquad (1)$$

where $M^{(k)}$ is the kth moment of the magnetization; $t$ is the time; $\tau$ is the reduced temperature; $L$ is the linear size of the system; $b$ is the scaling factor; $\beta$ and $\nu$ are static critical exponents of the magnetization and the correlation radius, respectively; $z$ is the dynamic critical exponent; and $x_0$ is a new independent critical exponent defining the scaling dimension of the initial magnetization $m_0$.

When starting from the low-temperature ordered state ($m_0 = 1$), the theory predicts the power behavior of the magnetization in the short-time mode at the critical point ($\tau = 0$) under the assumption $b = t^{1/z}$ in Eq. (1) for the systems with rather large linear sizes $L$:

$$M(t) \sim t^{-c_1}, \quad c_1 = \frac{\beta}{\nu z}. \tag{2}$$

By taking the logarithm of both sides of the equation and taking the derivative with respect to $\tau$ at $\tau = 0$, we obtain the power law for the logarithmic derivative:

$$\partial_\tau \ln M(t,\tau)\big|_{\tau=0} \sim t^{-c_{l1}}, \quad c_{l1} = \frac{1}{\nu z}. \tag{3}$$

For the Binder cumulant $U_L(t)$ calculated from the first and second moments of the magnetization, the finite-size scaling theory gives the following dependence at $\tau = 0$:

$$U_L(t) = \frac{M^{(2)}}{(M)^2} - 1 \sim t^{c_U}, \quad c_U = \frac{d}{z}. \tag{4}$$

Thus, in one numerical experiment, the short-time dynamics method makes it possible to determine the three critical exponents using Eqs. (2)–(4): $\beta$, $\nu$, and $z$. In addition, dependences (2) plotted for different temperatures make it possible to find the $T_c$ value from their deviation from the straight line in the log–log scale.

Using this method, we studied the critical relaxation from the low-temperature ordered state of the fully frustrated three dimensional Ising model on a simple cubic lattice. This model was proposed for the first time by Villain [8] in the two-dimensional case on a square lattice for the description of spin glasses. Further, it was generalized for the three dimensional case by Blankschtein [9]. This model is given schematically in Fig. 1.

This model is of interest because studies of frustrated systems are mainly focused on models on triangular and hexagonal lattices, while the properties of models on a cubic lattice were little studied. The dynamic critical behavior of such systems was also almost unstudied.

The Hamiltonian of the frustrated Ising model can be presented in the form

$$H = -\frac{1}{2} \sum_{<i,k>} J_{ik} S_i S_k, \quad S_i = \pm 1, \tag{5}$$

where $S_i$ is the Ising spin at the $i$th site of the lattice and $J_{ik}$ is the exchange interaction between spins for ferromagnetic ($J > 0$) and antiferromagnetic ($J < 0$) bonds. Frustrations in this model are due to the competition of exchange interactions [8].

We studied a cubic particle containing $L \times L \times L$ unit cells in each crystallographic direction with periodic boundary conditions. We considered a system with the linear size $L = 64$ containing $N = 262144$ spins. The $L$ value was chosen as the minimally necessary one in order to exclude the effect of the finite sizes on the result [1].

Calculations were performed by the standard Metropolis Monte Carlo algorithm. The relaxation of the system was performed from the initial completely ordered low-temperature state with the starting magnetization $m_0 = 1$ during the time $t_{max} = 1000$. One Monte Carlo step per spin was taken as the "time" unit. The relaxation dependences were calculated 50000 times. The obtained results were averaged between each other.

Critical temperatures were determined from the time dependence of the magnetization given by Eq. (2), which should be a straight line on the log–log scale at the phase transition point. The deviation from the straight line was determined by the least-squares method. The temperature at which this deviation was minimal was taken as the critical one. In the determination of $T_c$, the magnetization curves were analyzed with the step $\Delta T = 0.0001$ in units of the exchange integral $k_b T / |J|$.

The logarithmic derivative at the phase transition point was calculated by the least-squares approximation from three time dependences of the magnetization plotted for temperatures $T_c - 0.01$, $T_c$, and $T_c + 0.01$. In Fig. 2, these curves are shown on the log–log scale in the time interval $t = [100, 1000]$ (here and further, all quantities are given in arbitrary units). Figure 2 demonstrates vividly the temperature effect on the magnetization curve.

The time dependences of the magnetization, its logarithmic derivative, and the Binder cumulant obtained at the critical point are shown in Figs. 3–5, respectively, on the log–log scale in the time interval $t = [1, 1000]$. Points in the figures show the simulation results and solid lines show their least-squares approximation according to Eqs. (2)–(4). The analysis of the plots showed that the power scaling behavior of the studied system begins at a time instant of about $t = 100$. For this reason, the approximation of all curves was performed in the time interval $t = [200, 1000]$.

Our results for the critical temperature, static critical exponents of the magnetization and correlation radius, and the dynamic critical exponent are presented in the table in comparison with the results from [10–12], where the static critical properties of the fully frustrated Ising model were studied. The table demonstrates that our results for the critical temperature and static critical exponents are in good agreement with the results of these works. The dynamic critical exponent is close to that predicted theoretically for anisotropic magnets ($z = 2$, model A [6]). We note that the dynamic critical exponent for the studied model was obtained for the first time.

The results of this work demonstrate the efficiency of the application of the short-time dynamics method to studying the critical properties of three-dimensional models with frustration. The advantage of this method is that it provides not only the dynamic critical exponent but also the static critical exponents and critical temperature within one numerical experiment. In addition, the critical slowing down is not manifested in this approach, since the spatial correlation radius remains small in the short-time interval even near the critical point [7].

Critical temperature and critical exponents of the completely frustrated Ising model

| Parameter | This work | [10] | [11] | [12] |
|---|---|---|---|---|
| $T_c$ | 1.3487(1) | 1.344(2) | 1.355(2) | 1.347(1) |
| $\beta$ | 0.22(3) | 0.21(2) | – | 0.25(2) |
| $\nu$ | 0.54(3) | 0.55(2) | 0.55(2) | 0.56(2) |
| $z$ | 2.21(3) | – | – | – |

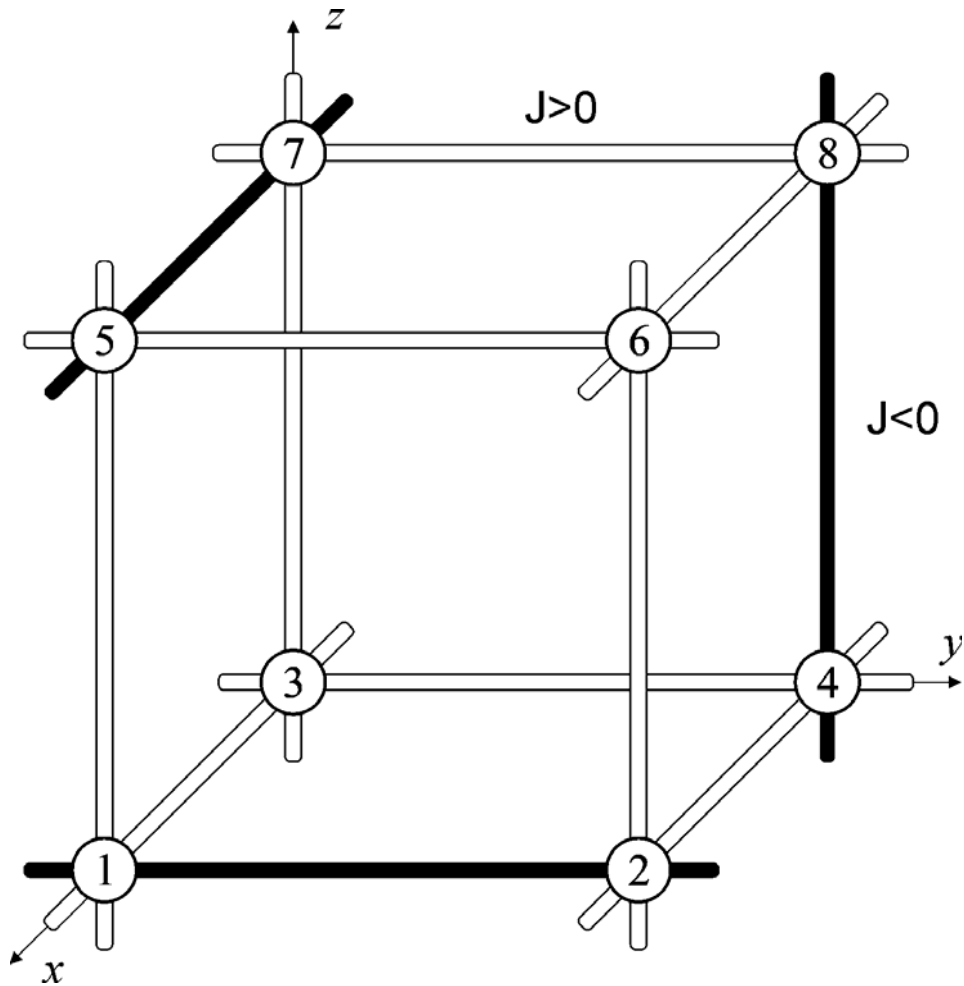

**Fig. 1**. Fully frustrated Ising model on a simple cubic lattice. White and black colors denote the ferromagnetic ($J > 0$) and antiferromagnetic ($J < 0$) bonds, respectively.

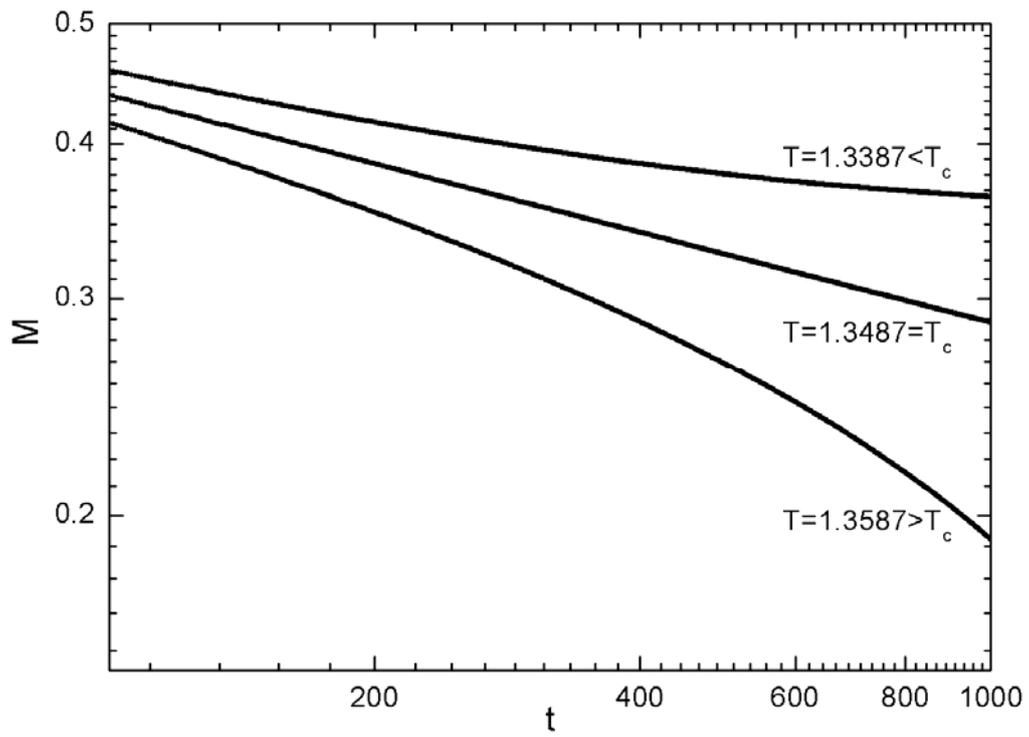

**Fig. 2.** Time dependence of the magnetization at three temperature values.

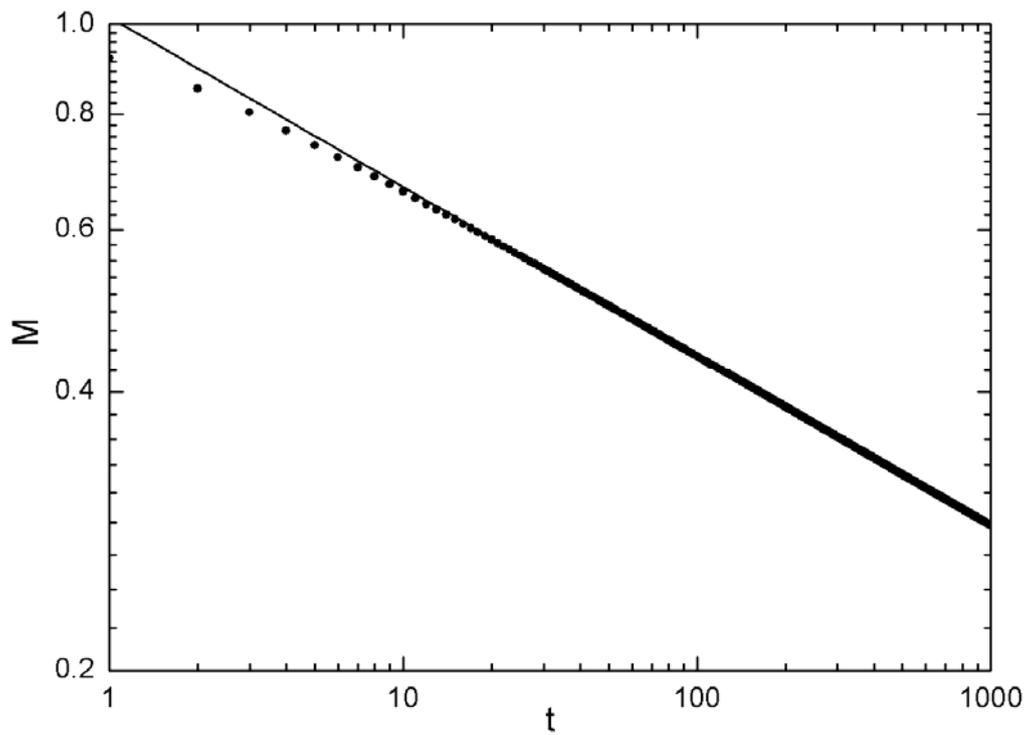

**Fig. 3.** Time dependence of the magnetization at the phase transition point.

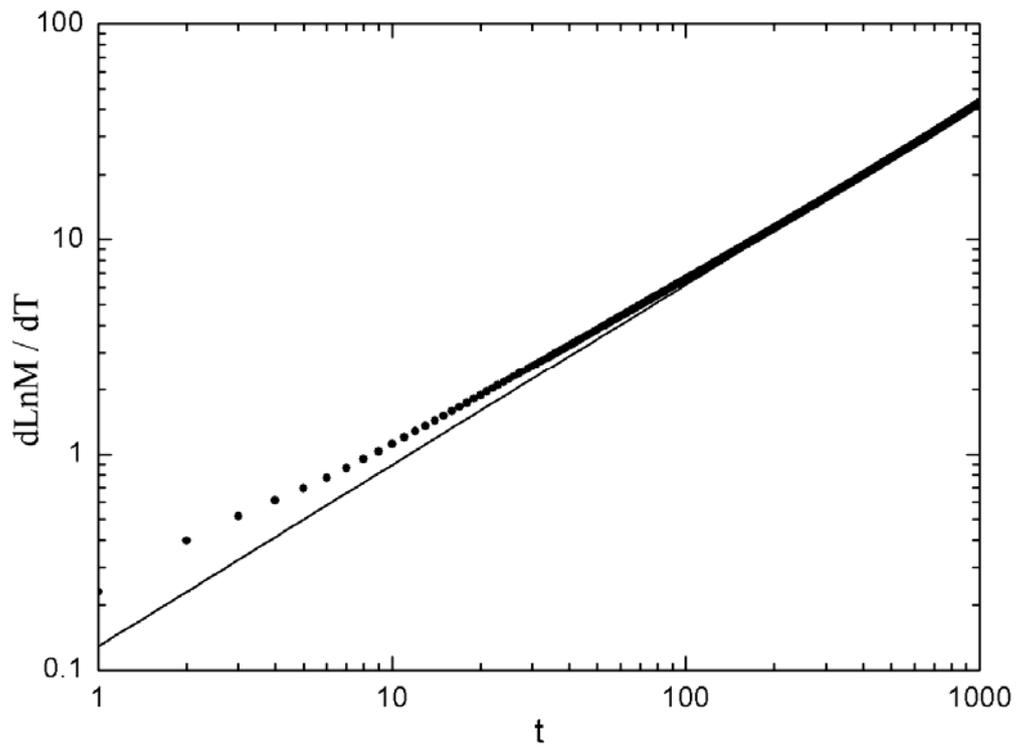

**Fig. 4.** Time dependence of the derivative of the logarithm of the magnetization at the phase transition point.

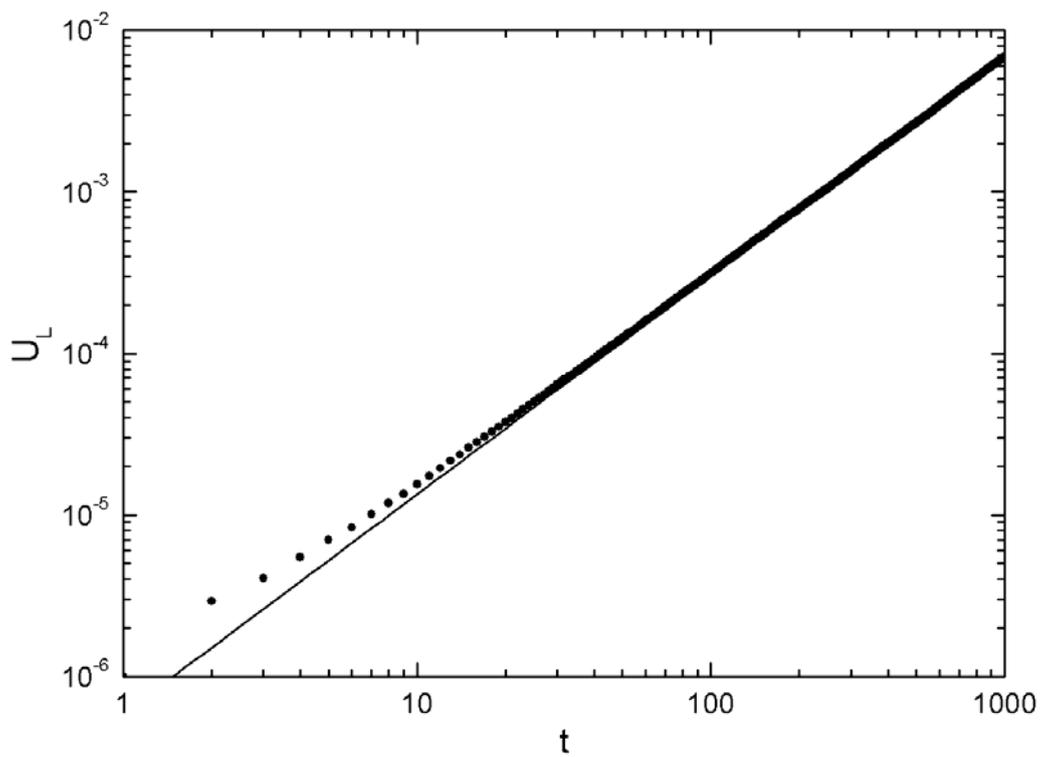

**Fig. 5.** Time dependence of the Binder cumulant at the phase transition point.